# Inverse design enables large-scale high-performance meta-optics reshaping virtual reality

Zhaoyi Li[1†], Raphaël Pestourie[2†], Joon-Suh Park[1,3], Yao-Wei Huang[1,4], Steven G. Johnson[2*], Federico Capasso[1*]

1. Harvard John A. Paulson School of Engineering and Applied Sciences, Harvard University, Cambridge, MA, USA.
2. Department of Mathematics, Massachusetts Institute of Technology, Cambridge, MA, USA.
3. Nanophotonics Research Center, Korea Institute of Science and Technology, Seoul, Republic of Korea.
4. Department of Electrical and Computer Engineering, National University of Singapore, Singapore.

† these authors contributed equally

* Corresponding emails: stevenj@math.mit.edu; capasso@seas.harvard.edu.

## Abstract

Meta-optics has achieved major breakthroughs in the past decade; however, conventional forward design faces challenges as functionality complexity and device size scale up. Inverse design aims at optimizing meta-optics design but has been currently limited by expensive brute-force numerical solvers to small devices, which are also difficult to realize experimentally. Here, we present a general inverse design framework for aperiodic large-scale complex meta-optics in three dimensions, which alleviates computational cost for both simulation and optimization via a fast-approximate solver and an adjoint method, respectively. Our framework naturally accounts for fabrication constraints via a surrogate model. In experiments, we demonstrate, for the first time, aberration-corrected metalenses working in the visible with high numerical aperture, poly-chromatic focusing, and large diameter up to centimeter scale. Such large-scale meta-optics opens a new paradigm for applications, and we demonstrate its potential for future virtual-reality platforms by using a meta-eyepiece and a laser back-illuminated micro-Liquid Crystal Display.

## Introduction

Meta-optics, a new class of planar optics, has reshaped the engineering of electromagnetic waves by using artificial subwavelength components or "meta-atoms" [1-6]. Recent breakthroughs in the physics [7-11] and advancements in large-scale meta-optics fabrication [12-14] inspire a vision for a future in which meta-optics will be widely used. Recent studies have demonstrated cutting-edge technologies based on meta-optics platforms, such as, polarization/light-field/depth imaging cameras [15-18], metasurface-driven OLEDs [19], virtual/augmented-reality systems [20,21], compact spectrometers [22-24], etc. So far, the mainstream design of the meta-optics is mostly based on a "forward" methodology, in which one engineers each individual meta-atom component (as a phase shifter) independently according to a predefined phase profile [25,26]. Forward design has demonstrated success in realizing simple device functions, like single-wavelength wave bending [27-29] or focusing [30,31]; however, it heavily relies on *a priori* intuitive knowledge and limits the development of large-scale complex meta-optics that can realize multiple custom functions depending on wavelengths, polarizations, spins, and angles of incident light. As the complexity, diameter, or constraints of a design problem scale up, the ability of a forward-driven method to search for an optimal solution becomes weaker and weaker. The future advancement of meta-optics demands a breakthrough in design philosophy.

In contrast to forward design, inverse design starts with desired functions and optimizes design geometries using computational algorithms. It has been a useful tool in solving large-scale complex engineering problems such as optimizing the shape of bridges or aircraft wings. In recent years, inverse design has been reshaping the landscape of photonics engineering. Multiple flavors of inverse design techniques have been studied: topological optimization techniques, which use a local gradient-based optimization tool to search for optimal photonic geometries [32,33]; machine-learning techniques [34-36], which train a neural network to find a design for a given response [37] or train a generative network (e.g., generative adversarial network) to sample the high-performance designs [38]. A recent evolution of inverse design in photonics optimizes the geometry and the post-processing parameters end-to-end [39-41]. Inverse design has demonstrated significant success in optimizing photonic crystals [42], on-chip nanophotonics [43,44], metasurfaces [45,46], and other devices.

Inverse design remains very challenging for aperiodic large-scale meta-optics. The optimization relies on many iterations of simulations, which become computationally intractable as design dimension scales up due to the multiscale nature of design problems [47]: the nanoscale meta-atom (nm) and the macroscale meta-optics (100s of μm to cm). For example, it is unrealistic to model an aperiodic 3D device with 1-cm diameter using the finite-difference time-domain (FDTD) or the finite element analysis (FEA) method, which can capture physics at nanoscale. On the other hand, ray-tracing simulations, which are suitable for large-scale optics design, cannot capture the full wave nature of the optical field. They also only allow slowly varying phase profiles, excluding the rich physics of rapidly varying phase wavefronts offered by engineered meta-atoms. To our knowledge, the diameter of inverse-designed fully three-dimensional metasurfaces has been limited to about 200λ [48-51], about 100 μm for visible light. In addition, most inverse-designed photonic geometries are hard to realize experimentally given realistic fabrication constraints [52].

In this paper, we present a generic inverse design framework that enables aperiodic large-scale three-dimensional complex-function meta-optics compatible with fabrication constraints. Our inverse-design method is computationally tractable (requiring a few hours on a desktop CPU) and advantageous for macroscale (1000s of λs) meta-optics design in tandem with exploitation of physics at the nanoscale. It greatly expands optical design to an unprecedented regime where conventional forward design is of limited use. These unique inverse design features enable experimental demonstration of meta-optics with high numerical aperture (NA = 0.7) and complex functionality. For example, we show polarization-insensitive RGB-achromatic metalenses and, even poly-chromatic metalenses. These inverse-designed meta-optics realizes, for the first time, mm- to cm-scale aperture size, which corresponds to an increase of four orders of magnitude in area compared with state of the art. To prove the potential of large-scale meta-optics in applications, we further demonstrate a meta-optics-based virtual-reality (VR) platform.

**Inverse design theoretical framework**

Fundamentally different from conventional forward design, the philosophy of inverse design is to start with the goal and then optimize it given the application's constraints. For the design of lenses, the goal is to maximize the intensity at the focal spot; that is, we max$imize\ I(\vec{x}_{target}, \vec{p})$ over a vector $\vec{p}$ of geometric parameters defining the metasurface, where $\vec{x}_{target}$ is the location of the focal spot [53]. For polychromatic lens design, the objective function is $max\left(\min_{\lambda \in \lambda_s}(I_\lambda(\vec{x}_{target}, \vec{p}))\right)$, where $\lambda_s$ is a discrete set of wavelengths of interest and $I_\lambda$ is the intensity function for a wavelength $\lambda$ [53]. This maximizes the focal intensity at multiple wavelengths simultaneously. We further reformulate this function to be differentiable as shown in the SI.

Fast and accurate "forward" evaluation of meta-optics performance is key to large-scale inverse design. We introduce a three dimensional (3D) fast approximate solver that is based on the convolution of local fields and Green's function (Fig. 1A). Accurate local fields above a training set of meta-atoms are computed in advance using rigorous coupled wave analysis (RCWA). A surrogate model, which is based on Chebyshev interpolation [54], is then built to fast predict the local field of an arbitrary meta-atom within the design regime bounded by fabrication constraints. By the equivalence principle, we convert the local fields to "artificial" sources of magnetic current density $\vec{S}_{local}(\vec{x}, \vec{p})$, and the focal intensity is computed by using a convolution between the current sources and vectorial Green's function [53]:

$$\left|\vec{E}(\vec{x}_{target})\right|^2 = \left|\int_\Sigma \vec{S}_{local}(\vec{x}, \vec{p}) \odot \overleftrightarrow{G}(\vec{x}, \vec{x}_{target}) d\vec{x}\right|^2 \text{ (Equation. 1)}$$

where $\vec{E}(\vec{x}_{target})$ is the electric field at the focal spot, $\odot$ represents the Hadamard product, and $\overleftrightarrow{G}(\vec{x}, \vec{x}_{target})$ is the dyadic Green's function from a local position $\vec{x}$ to a target position $\vec{x}_{target}$. Note that the Green's function only needs to be computed once and can be reused in subsequent optimization iterations. It is analytical in free space and does not require a paraxial approximation. Here, we use a local periodic approximation to predict the local fields, assuming neighboring meta-atoms are similar [53,55]. To further speed up our simulator, we impose cylindrical symmetry on the design parameters while retaining the tiling of the meta-atoms in Cartesian coordinates (see SI.). Unlike fully axis-symmetric designs [51], however, on a subwavelength scale our meta-atoms break cylindrical symmetry.

Optimization in a high-dimensional design space, when $\vec{p}$ is of dimension >> 1000, is another challenge for inverse design. Here, we use a local gradient-based optimization method, called a "conservative convex separable approximation" [56], to search for an optimal design consisting of $10^6$ to $10^9$ degrees of freedom. We also applied a multi-start approach by exploring multiple random initial design parameters [57]. For fast computation of the gradients $\nabla_\mathbf{p} I(\vec{x}_{target})$, we take advantage of an adjoint method [58], which can evaluate the gradients for all parameters simultaneously using only two simulations. In comparison, a traditional brute-force method needs (N+1) simulations, where N is the dimension of $\vec{p}$. The adjoint method is illustrated in Fig. 1A (details in the SI.):

$$\nabla_{\mathbf{p}}I(\vec{x}_{target}) = 2\Re(\int_{\Sigma}\left(\vec{E}(\vec{x}_{target})^{*}\nabla_{\mathbf{p}}\vec{S}_{local}(\vec{x},\vec{p})\right) \odot \vec{G}(\vec{x},\vec{x}_{target})d\vec{x}) \quad \text{(Equation. 2)}$$

where $\Re$ denotes the real part, $\vec{G}(\vec{x}, \vec{x}_{target})$ is the dyadic Green's function from a target position $\vec{x}_{target}$ to a local position $\vec{x}$, and $\nabla_{\mathbf{p}}\vec{S}_{local}(\vec{x},\vec{p})$ is the gradient of the local current source with respective to the design parameter $\vec{p}$, which can also be fast evaluated by using a pre-trained surrogate model at low cost. It means the gradient $\nabla_{\mathbf{p}}I(\vec{x}_{target})$ can be efficiently obtained everywhere at once in a backward simulation using an equivalent source $\left(\vec{E}(\vec{x}_{target})^{*}\nabla_{\mathbf{p}}\vec{S}_{local}(\vec{x},\vec{p})\right)$. The gradient information was then fed into the optimizer for meta-design update. The whole design flow is summarized in Fig. 1B. We started from a random meta-design and went through iterations of optimization loops, relying on a forward simulator and an adjoint simulator, until the device performance converged and met the design criteria. We then evaluated the final design in simulations and further in experiment.

**Large-scale inverse-designed polychromatic metalenses**

Engineering a focus at multiple wavelengths and in different polarization states simultaneously is challenging, especially in the case of high NA. By applying the inverse design method, we first demonstrated a polarization-insensitive, RGB-achromatic metalens. This metalens has a diameter of 2 mm and numerical aperture of 0.7. Figure 2A is an optical microscope image of the device fabricated using electron beam lithography and atomic layer deposition [59]. The inset scanning electron microscope image shows anisotropic $TiO_2$ nanofin structures with spatially varying inverse-designed geometries on top of a fused-silica substrate. The height of the nanofins is 600 nm and the square-lattice periodicity is 400 nm. Each nanofin has a rectangular shape whose sizes are determined by optimization and is aligned parallel to the unit-cell axes. It partially converts left-handed circularly polarized (LCP) light to right-handed circularly polarized (RCP) light, and vice versa (Fig. 1A). The polarization conversion from L ($|\sigma_-\rangle$) to R ($|\sigma_+\rangle$) and R to L is equal by symmetry in our case, as described by the Jones' matrix:

$$\begin{bmatrix}\tilde{E}_{LCP}^{out}\\\tilde{E}_{RCP}^{out}\end{bmatrix} = \begin{bmatrix}(\tilde{t}_L+\tilde{t}_s)/2 & (\tilde{t}_L+\tilde{t}_s)e^{-2i\alpha}/2\\(\tilde{t}_L-\tilde{t}_s)e^{+2i\alpha}/2 & (\tilde{t}_L+\tilde{t}_s)/2\end{bmatrix}\begin{bmatrix}\tilde{E}_{LCP}^{in}\\\tilde{E}_{RCP}^{in}\end{bmatrix} \quad \text{(Equation. 3)}$$

Where $\tilde{t}_L$ and $\tilde{t}_s$ are complex transmission along long and short axis, respectively, $\alpha$ is the rotation angle of nanofin, "out" means output field, and "in" means input field. Due to this symmetry and the fact that any polarization state can be written as superposition of LCP and RCP fields, our metalens design can focus light equally well for any arbitrary polarization state [59,60]. Figure 2B shows the simulation results for the focal intensity distribution along the optical axis at the design RGB wavelengths of 488 nm, 532 nm, and 658 nm. These wavelengths are chosen to correspond to our single-wavelength laser diodes. The inset is the zoomed-in view of the focal peaks, which shows achromatic focusing with negligible focal shifts (< 50 nm). Figure 2C is the measured focal intensity distribution at the RGB wavelengths in the XZ plane, where X is along the lens radial direction and Z is along the optical axis. The maximum focal shift is ~ 1.5 μm, which is ~ 0.15% of the focal length. Figure 2D, from top to bottom, is the measured intensity distribution at the focal planes of the blue, green, and red wavelengths, respectively.

Their respective measured focal intensity profiles, implying diffraction-limited focus (Detailed analysis can be found in the SI.) We measured the absolute focusing efficiency, which is defined as the ratio between the power in the focal spots and the incident power, as a function of the incidence polarization angles. Figure 2F shows that the absolute efficiency is about 15% at RGB wavelengths and is independent of the polarization angle of the incident light. Moreover, our metalens focuses light of an arbitrary polarization state to its orthogonal state, which is useful for improving the imaging contrast. We further characterized the imaging performance of the metalens using the United States Air Force (USAF) resolution target. Figure 2, G to I, is the imaging result of the element No. 5 and No. 6 from group No. 7 under blue, green, and red illumination. The smallest feature size is 2.2 µm and can be clearly resolved. To demonstrate achromatic imaging, we further imaged the same area using synthesized white-light illumination by mixing RGB color in the incident light. The result is a clear whitish image with the same magnification (Fig. 1J). More imaging results under other synthesized light illumination can be found in SI.

The inverse-design method has more pronounced advantages over conventional forward design methods when designing a meta-optics with more complicated functions. Forward-design methods, like wavefront phase matching, struggle in the regime where no single meta-atom can simultaneously satisfy the targeted phase profiles for multiple functions. They also neglect the effect of a non-uniform amplitude or phase profile. A good phase matching sometimes comes at the cost of low efficiency due to the intrinsic correlation between the phase and amplitude of the engineered electromagnetic wave by meta-atoms. Moreover, forward designs are usually one-way without feedback loops, and thus do not provide confirmation of optimality or robustness. Importantly, forward designs require *a priori* knowledge of the desired wave solution, which is unavailable for complex problems. In comparison, our inverse-design method can obtain previously unknown solutions to complex design problems, because it starts only with the design objective and iteratively searches for an optimal solution in a hyperdimensional design space.

To prove the concept, we further demonstrated the first experimental polychromatic metalenses with six-wavelength-achromatic focusing performance for visible light. These two metalenses have an aperture diameter of 2 mm and NAs of 0.3 and 0.7. Figure 3A is the SEM image of the NA = 0.3 metalens. This metalens is designed for achromatic focusing at six wavelengths of 490 nm, 520 nm, 540 nm, 570 nm, 610 nm, and 650 nm. Figure 3B is the simulation result showing their focal intensity distribution along the optical axis. The measured focusing intensity (Fig. 3C) in the XZ plane shows good agreement with the simulation results (Fig. 3C). The maximum focal shift among design wavelengths is 500 nm (< 0.02% of the focal length). The simulation and measurement results of the NA = 0.7 metalens are shown in the SI. Figure 3D shows that the measured full-width-half-maximums (FWHMs) of the focal spots in comparison with the ideal Airy-function theory. The subtle differences are because we used a super-continuum laser as the light source, which has a larger linewidth (FMHWs) of ~5 nm in comparison with ~0.5 nm linewidth of laser diodes (SI.).  Figure 3, E to J, is the measured focal intensity distribution at a common focal plane of six design wavelengths. The measurement results of the NA = 0.7 metalens are shown in the SI.

To further prove the scalability of our inverse design method, we designed and fabricated a cm-scale metalens. This metalens is designed for achromatic focusing at RGB wavelengths with an NA of 0.3. Figure 4A shows the 1-cm-diameter RGB-achromatic flat meta-optics on 2-inch fused silica wafer with a reference ruler behind. The inset is the SEM image showing the meta-atoms building blocks. We utilized a fast E-beam writer and operated at a high current. In this way, we achieved 10-nm structural resolution at a low cost in fabrication time. Figure 4B is the simulation result showing the focal intensity distribution along the optical axis at design wavelengths, and the inset is the zoomed-in view of the peaks to show its achromatic focusing performance. The measured focal intensity distribution in the XZ plane is shown in Fig. 4C. The maximum focal shift among RGB wavelengths is ~ 4.5 µm, which is ~ 0.03% of the design focal length. Figure 4, D to F, shows the measured focal intensity distribution at the focal planes of $\lambda$ = 488 nm, 532 nm, and 658 nm. The slight distortion of the focal spots is due to the non-uniform incident illumination over the cm-scale lens' aperture. We further characterized the metalens by imaging the whole group No. 7 of the USAF resolution targets. Figure 4, G to H, is the imaging result under illumination of blue, green, and red incident light, respectively, which shows excellent imaging performance.

**Virtual-reality imaging demonstration**

Large-scale meta-optics may have significant impact on many applications. Here, we demonstrate a virtual-reality (VR) imaging system based on our meta-optics. VR is a technology that creates an immersive experience by replacing reality with an imaginary world [61]. Its recent breakthroughs have not only attracted attention from the scientific community and industry but have also piqued the interest of the general public. Unfortunately, widespread use of VR devices has been hindered by a bottleneck in the optical architecture. The eyepieces used in current VR headsets mostly rely on refractive singlets, which suffer from bulky size and weight, and they furthermore compromise the viewing experience due to spherical and chromatic aberrations [62]. Meta-optics offer a technology to address these challenges of current VR systems [21].

Figure 5A is the schematic of our VR system, based on our cm-scale RGB-achromatic meta-eyepiece and a laser-illuminated micro-LCD. The micro-LCD is placed close to the focal plane of the meta-eyepiece, and the image on the display is projected via the meta-eyepiece onto the retina, creating a virtual scene. In the experiment, we used a tube lens to mimic the cornea and eye crystalline lens and a CMOS camera to mimic the retina. In addition, we home-built a near-eye display using the laser light as the back-illumination source. Such a display offers high brightness and a wide color gamut due to the narrow linewidth. The pixel size is about 8 µm, matching the state of the art. The key components of the meta-eyepiece and the display are illustrated in the dashed brown box of Figure 5B. We first demonstrate binary VR imaging. Figure 5C shows the VR image of a red letter-H shield logo, and Figure 5D is the zoomed-in view of one corner (from the white dashed box of Figure 5C). One can see that the meta-eyepiece resolves every pixel of the display. Figure 5, E and F is the imaging result for an MIT logo under green and blue illumination, respectively. We further demonstrated grayscale VR imaging. Figure 5, G and H is a grayscale imaging result (in red light) showing a Harvard building and statue, respectively. Figure 5, I and H, shows the grayscale VR images of a building and lighthouse in green and blue, respectively. These RGB-color imaging results imply an ability to image in full color, because color images are simply formed by mixing these primary colors. For example, figure 5, K to M, shows VR imaging of distinct red, green, and blue circles, respectively. Figure 5N is the simulated color VR imaging result by superimposing Fig.5, K to M, which show synthesized colors of yellow, magenta, cyan, and white in the circle overlapping regions. Furthermore, figure 5, O to Q, shows VR imaging of a Harvard tower in red, green, and blue channel. Figure 5R is the simulated full-color imaging result by superimposing the RGB images (Fig. 5S to 5V). In addition to the static VR images, our VR system can also display a dynamic VR object. Figure 5, O to R, displays a running cat that is captured at 0 ms, 180 ms, 460 ms, and 600 ms, respectively. The near-eye display has a refresh rate of 60 Hz, and the recorded movie can be found in the SI. We further discussed a strategy to reduce form factor of the display by using a metasurface-based illumination plate (SI.).

This work shows major advances over the previous VR system [21]. Thanks to the innovative inverse design method, the meta-optics has increased the aperture size from mm to cm, which means it can be integrated with micro displays and is more realistic for applications. Micro displays are future trend for VR optical engines; however, there has not yet been an eyepiece solution that can resolve high-resolution (~5 μm) color images. Second, the meta-optics now has polarization-insensitive focusing performance, which alleviates additional polarization-selection components (e.g., linear polarizer and phase retarder) and makes better use of incident light. Third, the meta-atoms now have a simple geometry shape and, thus, are more compatible with large-scale and mass production. Finally, displaying a movie is now possible thanks to the high refresh rate of our system. In the future, we believe meta-optics will augment conventional lens platforms [63] to form a high-performance aberration-free compact hybrid eyepiece for VR/AR.

**Discussion**

In this paper, we presented a general inverse design framework that is suitable for the large-scale 3D photonic-device optimization. We demonstrated, for the first time, inverse-designed 3D meta-optics of large diameter, including 2-mm-diameter RGB-achromatic and polychromatic metalenses, and even a cm-scale RGB-achromatic metalens, the largest to date, which consists of ~ $10^{12}$ meta-atoms. Furthermore, we demonstrate a path towards a future virtual-reality platform based on a meta-eyepiece and a laser-illuminated micro-LCD. This inverse design method is also applicable to optimizing other optical elements in a VR/AR system, such as optical combiners.

For further performance improvements, one could design freeform and multilayered meta-optics by adding many more degrees of freedom to the unit cells. In addition, the Chebyshev interpolation surrogate model used in this work needs an exponentially increasing dataset for more design parameters, but recent studies indicate that active-learning neural-network techniques can more efficiently model meta-atoms with more than 4 design parameters and multiple layers [64]. Fully freeform topology optimization has also begun to steadily approach larger scales by exploiting domain-decomposition approximations [48,65] and axisymmetric restrictions [51] with the help of large-scale computing power. Besides metalens optimization, one can also take advantage of inverse design to explore additional physical processes, such as nonlinear effects, and to gain better understating of the multi-functional trade-offs in photonic platforms. We believe that a more and more important role will be played by large-scale inverse-design methods in the future development of meta-optics.

## Methods

Simulation: The meta-atoms are simulated using the method of rigorous coupled wave analysis (RCWA). In the simulation setup, the height of the $TiO_2$ meta-atoms is 600 nm, the periodicity of the unit cell is 400 nm, and the substrate is fused silica. The incident light is configured to LCP (RCP), and monitored light is in the opposite polarization state of RCP (LCP). The simulation wavelength sweeps from 480 nm to 680 nm in the visible.

Fabrication: The metalenses are fabricated on glass wafers. The fabrication starts with spin coating of resists in the following manner: an atomic-thin layer of HMDS, a layer of 600-nm-thick ZEP (EBR, Zeon Chemicals, ZEP-520A), and then a final layer of conductive polymer (Showa Denko, ESPACER 300) to dissipate charges during the following electron beam lithography (EBL). After that, the sample was exposed using Elionix, ELS-F12 followed by development in water and then in o-xylene solution to remove the conductive polymer and the exposed ZEP resist, respectively. Next, a thin film of ~210-nm-thick $TiO_2$ was deposited onto the developed sample using atomic layer deposition (Cambridge, Nanotech, Savannah). $TiO_2$ thin film uniformly coated the sample not only inside the exposed area but also on top of the unexposed resist. The excessive $TiO_2$ layer was later removed using reactive ion etching (Oxford Instrument, PlasmaPro 100 Cobra 300) with etchant gases of $CHF_3$, O2, and Ar. In the final step, the etched sample was lifted-off in solution of remover PG (MicroChem Corporation) at 85°C for 24hrs to strip off the resist.

## Acknowledgements

Funding: This work was supported by the Defense Advanced Research Projects Agency (grant# HR00111810001). This work was performed in part at the Center for Nanoscale System (CNS), a member of the National Nanotechnology Coordinated Infrastructure (NNCI), which is supported by the National Science Foundation under NSF award no. 1541959. CNS is part of Harvard University. R. P. was supported by the U.S. Army Research Office through the Institute for Soldier Nanotechnologies (Award No. W911NF-18-2-0048) and the MIT- IBM Watson AI Laboratory (Challenge No. 2415). Y.-W.H. is supported by the National Research Foundation, Prime Minister's Office, Singapore under its Competitive Research Program (CRP Award No. NRF-CRP15-2015-03).

## Author contributions

Z.L. and R.P. conceived the original concept. R.P. developed the inverse design framework with contribution from Z.L. Z.L. conducted the metalenses fabrication, measurement, and virtual reality imaging experiment. J.P. and Y.H. contributed to the device fabrication and SEM imaging. S.J. and F.C. supervised the project. Z.L. and R.P. prepared the manuscript with input from the authors.

## Competing interests

The authors declare no competing interests.

## References

1. Yu, N. & Capasso, F. Flat optics with designer metasurfaces. *Nature materials* **13**, 139-150 (2014).
2. Chen, H.-T., Taylor, A. J. & Yu, N. A review of metasurfaces: physics and applications. *Reports on progress in physics* **79**, 076401 (2016).
3. Khorasaninejad, M. & Capasso, F. Metalenses: Versatile multifunctional photonic components. *Science* **358** (2017).
4. Lalanne, P. & Chavel, P. Metalenses at visible wavelengths: past, present, perspectives. *Laser & Photonics Reviews* **11**, 1600295 (2017).
5. Tseng, M. L. *et al.* Metalenses: advances and applications. *Advanced Optical Materials* **6**, 1800554 (2018).
6. He, Q., Sun, S., Xiao, S. & Zhou, L. High-efficiency metasurfaces: principles, realizations, and applications. *Advanced Optical Materials* **6**, 1800415 (2018).
7. Wang, S. *et al.* Broadband achromatic optical metasurface devices. *Nature communications* **8**, 1-9 (2017).
8. Shrestha, S., Overvig, A. C., Lu, M., Stein, A. & Yu, N. Broadband achromatic dielectric metalenses. *Light: Science & Applications* **7**, 1-11 (2018).
9. Wang, S. *et al.* A broadband achromatic metalens in the visible. *Nature nanotechnology* **13**, 227-232 (2018).
10. Chen, W. T. *et al.* A broadband achromatic metalens for focusing and imaging in the visible. *Nature nanotechnology* **13**, 220-226 (2018).
11. Chen, W. T., Zhu, A. Y. & Capasso, F. Flat optics with dispersion-engineered metasurfaces. *Nature Reviews Materials* **5**, 604-620 (2020).
12. She, A., Zhang, S., Shian, S., Clarke, D. R. & Capasso, F. Large area metalenses: design, characterization, and mass manufacturing. *Optics Express* **26**, 1573-1585 (2018).
13. Park, J.-S. *et al.* All-glass, large metalens at visible wavelength using deep-ultraviolet projection lithography. *Nano letters* **19**, 8673-8682 (2019).
14. Hu, T. *et al.* Demonstration of a-Si metalenses on a 12-inch glass wafer by CMOS-compatible technology. *arXiv preprint arXiv:1906.11764* (2019).
15. Camayd-Muñoz, P., Ballew, C., Roberts, G. & Faraon, A. Multifunctional volumetric meta-optics for color and polarization image sensors. *Optica* **7**, 280-283 (2020).
16. Rubin, N. A. *et al.* Matrix Fourier optics enables a compact full-Stokes polarization camera. *Science* **365** (2019).
17. Lin, R. J. *et al.* Achromatic metalens array for full-colour light-field imaging. *Nature nanotechnology* **14**, 227-231 (2019).
18. Guo, Q. *et al.* Compact single-shot metalens depth sensors inspired by eyes of jumping spiders. *Proceedings of the National Academy of Sciences* **116**, 22959-22965 (2019).
19. Joo, W.-J. *et al.* Metasurface-driven OLED displays beyond 10,000 pixels per inch. *Science* **370**, 459-463 (2020).
20. Lee, G.-Y. *et al.* Metasurface eyepiece for augmented reality. *Nature communications* **9**, 1-10 (2018).
21. Li, Z. *et al.* Meta-optics achieves RGB-achromatic focusing for virtual reality. *Science Advances* **7**, eabe4458 (2021).
22. Faraji-Dana, M. *et al.* Compact folded metasurface spectrometer. *Nature communications* **9**, 1-8 (2018).
23. Zhu, A. Y. *et al.* Compact aberration-corrected spectrometers in the visible using dispersion-tailored metasurfaces. *Advanced Optical Materials* **7**, 1801144 (2019).

24. McClung, A., Samudrala, S., Torfeh, M., Mansouree, M. & Arbabi, A. Snapshot spectral imaging with parallel metasystems. *Science advances* **6**, eabc7646 (2020).
25. Lalanne, P. Waveguiding in blazed-binary diffractive elements. *JOSA A* **16**, 2517-2520 (1999).
26. Khorasaninejad, M. & Capasso, F. Broadband multifunctional efficient meta-gratings based on dielectric waveguide phase shifters. *Nano letters* **15**, 6709-6715 (2015).
27. Zhang, L. *et al.* Ultra-thin high-efficiency mid-infrared transmissive Huygens meta-optics. *Nature communications* **9**, 1-9 (2018).
28. Yu, N. *et al.* Light propagation with phase discontinuities: generalized laws of reflection and refraction. *science* **334**, 333-337 (2011).
29. Ni, X., Emani, N. K., Kildishev, A. V., Boltasseva, A. & Shalaev, V. M. Broadband light bending with plasmonic nanoantennas. *Science* **335**, 427-427 (2012).
30. Aieta, F. *et al.* Aberration-free ultrathin flat lenses and axicons at telecom wavelengths based on plasmonic metasurfaces. *Nano letters* **12**, 4932-4936 (2012).
31. Khorasaninejad, M. *et al.* Metalenses at visible wavelengths: Diffraction-limited focusing and subwavelength resolution imaging. *Science* **352**, 1190-1194 (2016).
32. Molesky, S. *et al.* Inverse design in nanophotonics. *Nature Photonics* **12**, 659-670 (2018).
33. Sell, D., Yang, J., Doshay, S., Yang, R. & Fan, J. A. Large-angle, multifunctional metagratings based on freeform multimode geometries. *Nano letters* **17**, 3752-3757 (2017).
34. Ma, W. *et al.* Deep learning for the design of photonic structures. *Nature Photonics*, 1-14 (2020).
35. Trivedi, R., Su, L., Lu, J., Schubert, M. F. & Vuckovic, J. Data-driven acceleration of photonic simulations. *Scientific reports* **9**, 1-7 (2019).
36. Lu, L. *et al.* Physics-informed neural networks with hard constraints for inverse design. *axXiv* (2021).
37. An, S. *et al.* A deep learning approach for objective-driven all-dielectric metasurface design. *ACS Photonics* **6**, 3196-3207 (2019).
38. Jiang, J., Chen, M. & Fan, J. A. Deep neural networks for the evaluation and design of photonic devices. *Nature Reviews Materials*, 1-22 (2020).
39. Lin, Z. *et al.* End-to-end nanophotonic inverse design for imaging and polarimetry. *Nanophotonics* **1** (2020).
40. Tseng, E. *et al.* Neural Nano-Optics for High-quality Thin Lens Imaging. *arXiv preprint arXiv:2102.11579* (2021).
41. Sitzmann, V. *et al.* End-to-end optimization of optics and image processing for achromatic extended depth of field and super-resolution imaging. *ACM Transactions on Graphics (TOG)* **37**, 1-13 (2018).
42. Dory, C. *et al.* Inverse-designed diamond photonics. *Nature communications* **10**, 1-7 (2019).
43. Sapra, N. V. *et al.* On-chip integrated laser-driven particle accelerator. *Science* **367**, 79-83 (2020).
44. Piggott, A. Y. *et al.* Inverse design and demonstration of a compact and broadband on-chip wavelength demultiplexer. *Nature Photonics* **9**, 374-377 (2015).
45. Campbell, S. D. *et al.* Review of numerical optimization techniques for meta-device design. *Optical Materials Express* **9**, 1842-1863 (2019).
46. Chung, H. & Miller, O. D. High-NA achromatic metalenses by inverse design. *Optics Express* **28**, 6945-6965 (2020).
47. Backer, A. S. Computational inverse design for cascaded systems of metasurface optics. *Optics Express* **27**, 30308-30331 (2019).
48. Phan, T. *et al.* High-efficiency, large-area, topology-optimized metasurfaces. *Light: Science & Applications* **8**, 1-9 (2019).
49. Bayati, E. *et al.* Inverse designed metalenses with extended depth of focus. *ACS photonics* **7**, 873-878 (2020).
50. Mansouree, M. *et al.* Multifunctional 2.5 d metastructures enabled by adjoint optimization. *Optica* **7**, 77-84 (2020).

51. Christiansen, R. E. *et al.* Fullwave Maxwell inverse design of axisymmetric, tunable, and multi-scale multi-wavelength metalenses. *Optics Express* **28**, 33854-33868 (2020).
52. Vercruysse, D., Sapra, N. V., Su, L., Trivedi, R. & Vučković, J. Analytical level set fabrication constraints for inverse design. *Scientific reports* **9**, 1-7 (2019).
53. Pestourie, R. *et al.* Inverse design of large-area metasurfaces. *Optics Express* **26**, 33732-33747 (2018).
54. Boyd, J. P. *Chebyshev and Fourier spectral methods*. (Courier Corporation, 2001).
55. Pérez-Arancibia, C., Pestourie, R. & Johnson, S. G. Sideways adiabaticity: beyond ray optics for slowly varying metasurfaces. *Optics Express* **26**, 30202-30230 (2018).
56. Svanberg, K. A class of globally convergent optimization methods based on conservative convex separable approximations. *SIAM Journal on Optimization* **12**, 555-573 (2002).
57. Kan, A. R. & Timmer, G. T. Stochastic global optimization methods part I: Clustering methods. *Mathematical programming* **39**, 27-56 (1987).
58. Strang, G. *Computational science and engineering*. (2007).
59. Devlin, R. C., Khorasaninejad, M., Chen, W. T., Oh, J. & Capasso, F. Broadband high-efficiency dielectric metasurfaces for the visible spectrum. *Proceedings of the National Academy of Sciences* **113**, 10473-10478 (2016).
60. Chen, W. T., Zhu, A. Y., Sisler, J., Bharwani, Z. & Capasso, F. A broadband achromatic polarization-insensitive metalens consisting of anisotropic nanostructures. *Nature communications* **10**, 1-7 (2019).
61. Sherman, W. R. & Craig, A. B. *Understanding virtual reality: Interface, application, and design*. (Morgan Kaufmann, 2018).
62. Beams, R., Kim, A. S. & Badano, A. Transverse chromatic aberration in virtual reality head-mounted displays. *Optics Express* **27**, 24877-24884 (2019).
63. Chen, W. T. *et al.* Broadband achromatic metasurface-refractive optics. *Nano letters* **18**, 7801-7808 (2018).
64. Pestourie, R., Mroueh, Y., Nguyen, T. V., Das, P. & Johnson, S. G. Active learning of deep surrogates for PDEs: Application to metasurface design. *npj Computational Materials* **6**, 1-7 (2020).
65. Lin, Z. & Johnson, S. G. Overlapping domains for topology optimization of large-area metasurfaces. *Optics Express* **27**, 32445-32453 (2019).

**(A)**

### Forward simulator

**Fast approximate solver**

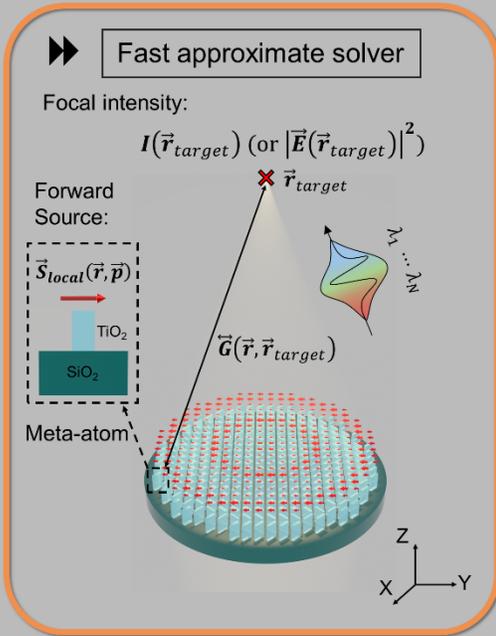

### Adjoint method

**Backward simulation**

**Gradient map computation**

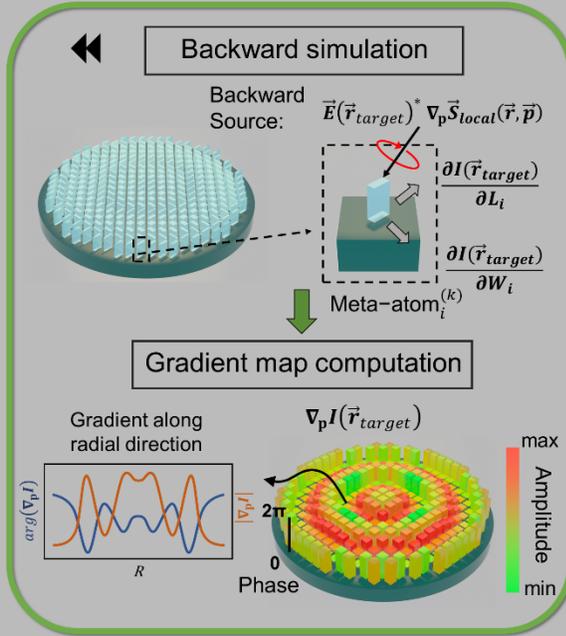

### Polarization conversion

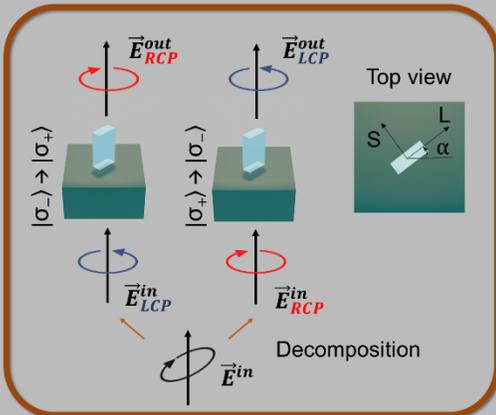

### Gradient-based Optimization

**Meta-design update**

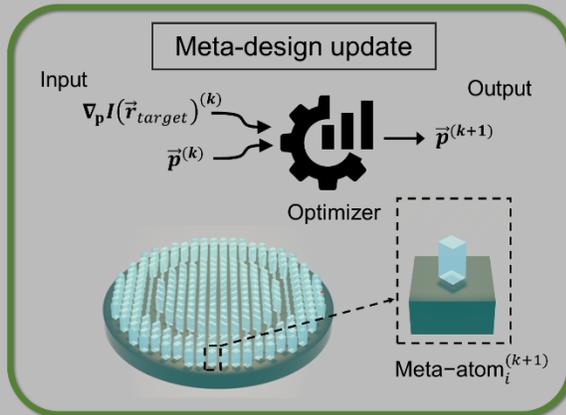

**(B)**

### Design flow chart

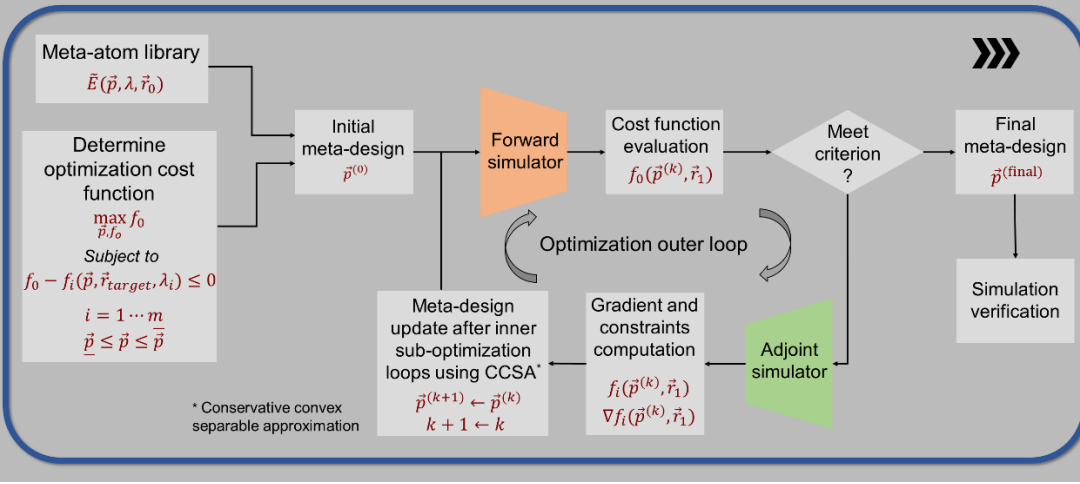

**Figure 1. Meta-optics inverse-design framework and flow chart.** (A), Bases of the inverse-design framework: 1. forward simulator by fast approximate solver, which evaluates the intensity of the field at the target via a convolution of the equivalent current with the appropriate Green's function; 2, polarization conversion by meta-atoms as described by Jones' matrix; 3. adjoint method that computes the gradient with respect to all the design parameters of the metasurface at a cost of a single simulation with a customized backward source; 4. a gradient-based optimization method, which updates the metasurface design through iterations. (B), Flow chart of the large-scale, poly-achromatic meta-optics design. With prior knowledge of the meta-atom library and optimization problem, we start with a random metasurface design and then update the design through optimization loops that consists of a forward simulator and an adjoint-based optimization engine. Once the criterion is met, we terminate the design loop and validate the design in simulation.

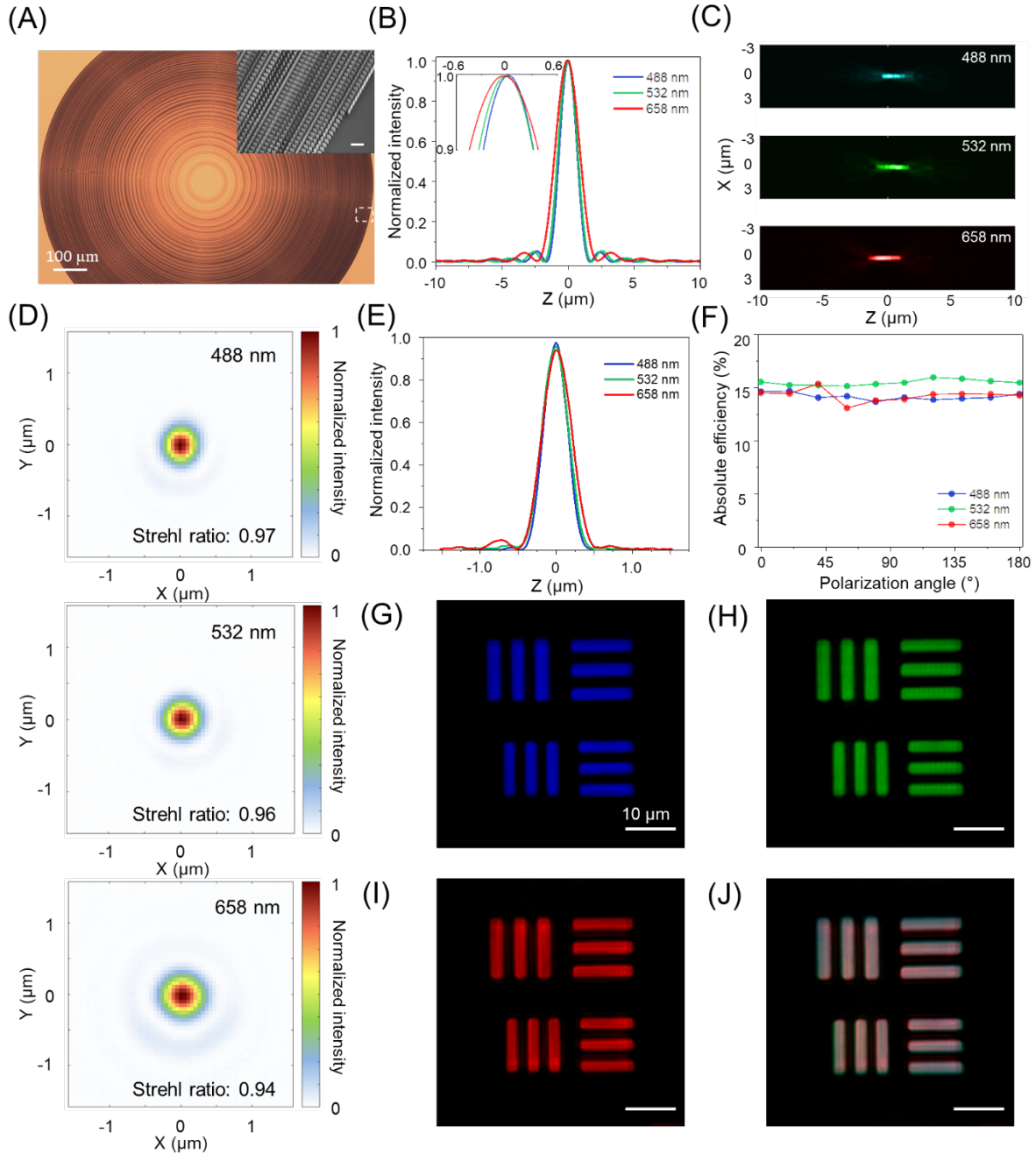

**Figure 2. 2-mm-diameter RGB-achromatic polarization-insensitive metalens (NA = 0.7)** (A), Optical microscope image of the fabricated device. Scale bar is 100 μm. The inset is the SEM image that corresponds to the region within the white dashed box. Scale bar is 1 μm. (B), Simulations of the normalized focal intensity along the optical axis (Z) at the three design RGB wavelengths. The inset shows the zoomed-in view of the intensity peaks. (C), Measured focal intensity distribution in the XZ plane. (D), Measured focal plane intensity distribution at λ = 488 nm, 532 nm, and 658 nm (from top to bottom, respectively) (E), Measured focal intensity profile of RGB focal spots. The peak values are normalized according to Airy functions (SI.). (F), Measured focusing efficiency as function of polarization angle of the incident light, showing polarization-insensitive focusing. (G)- (J), Imaging

results of the element #5 and #6 from the group #7 of the USAF resolution target at λ = 488 nm, 532 nm, and 658 nm, respectively. The scale bar is 10 µm unless noted. (J) Imaging result under synthesized white light illumination using RGB light.

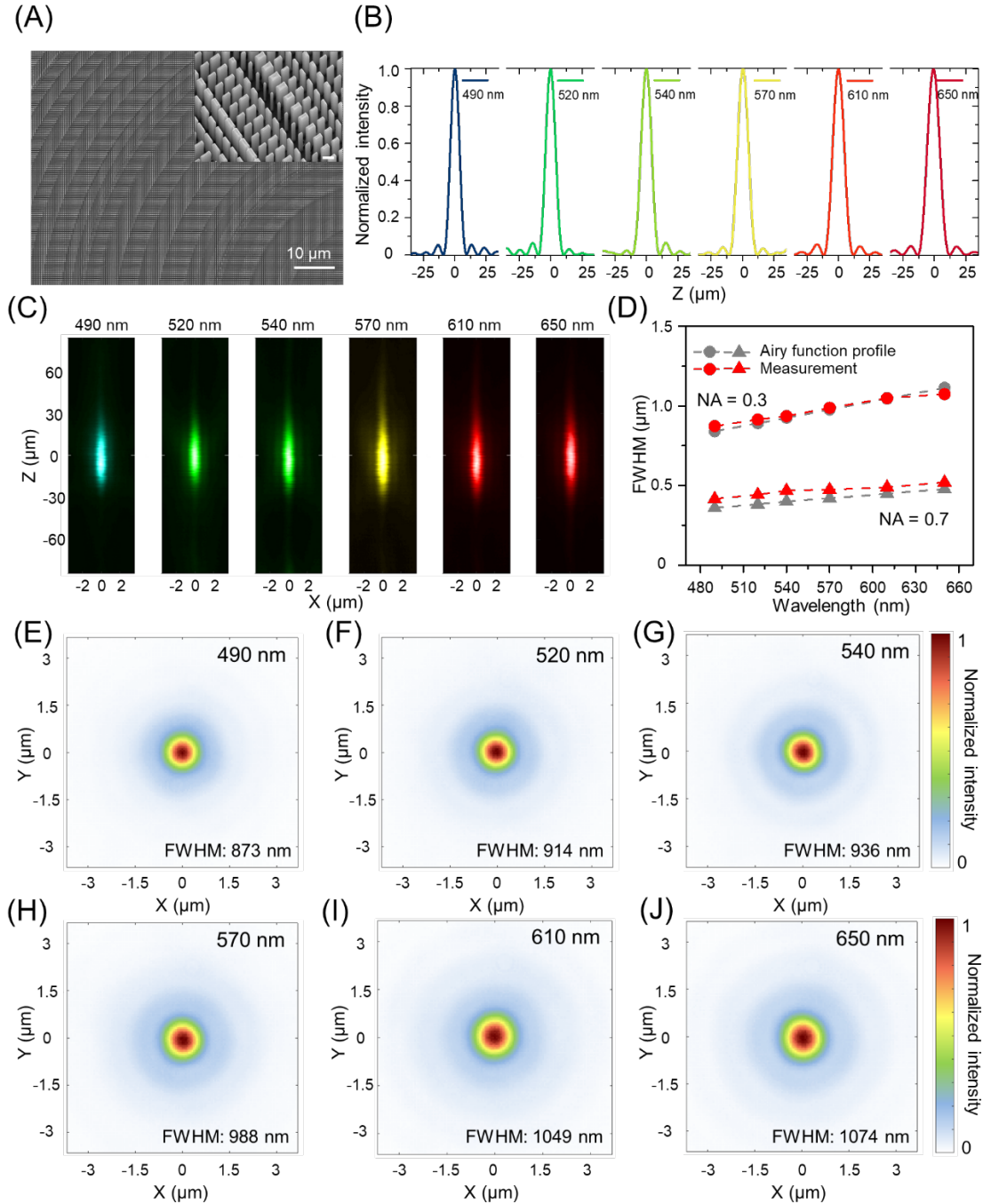

**Figure 3. 2-mm-diameter poly-chromatic polarization-insensitive metalenses (NA = 0.3 and 0.7)** (A), SEM image of the fabricated metalens with NA = 0.3. The scale bar is 10 µm. The inset is a zoomed-in tilted view. The scale bar is 400 nm. (B), Simulations of normalized focal intensity distribution along the optical axis at six design wavelengths (NA = 0.3). (C), Measured focal intensity distribution in the XZ plane at six wavelengths (NA = 0.3). (D), Measured full-width-half-maximum (FWHMs) of the focal

spots at six design wavelengths in comparison with ideal Airy function profile (NA = 0.3 and 0.7). (E) – (J), Measured focal plane intensity distribution at the design wavelengths (NA = 0.3).

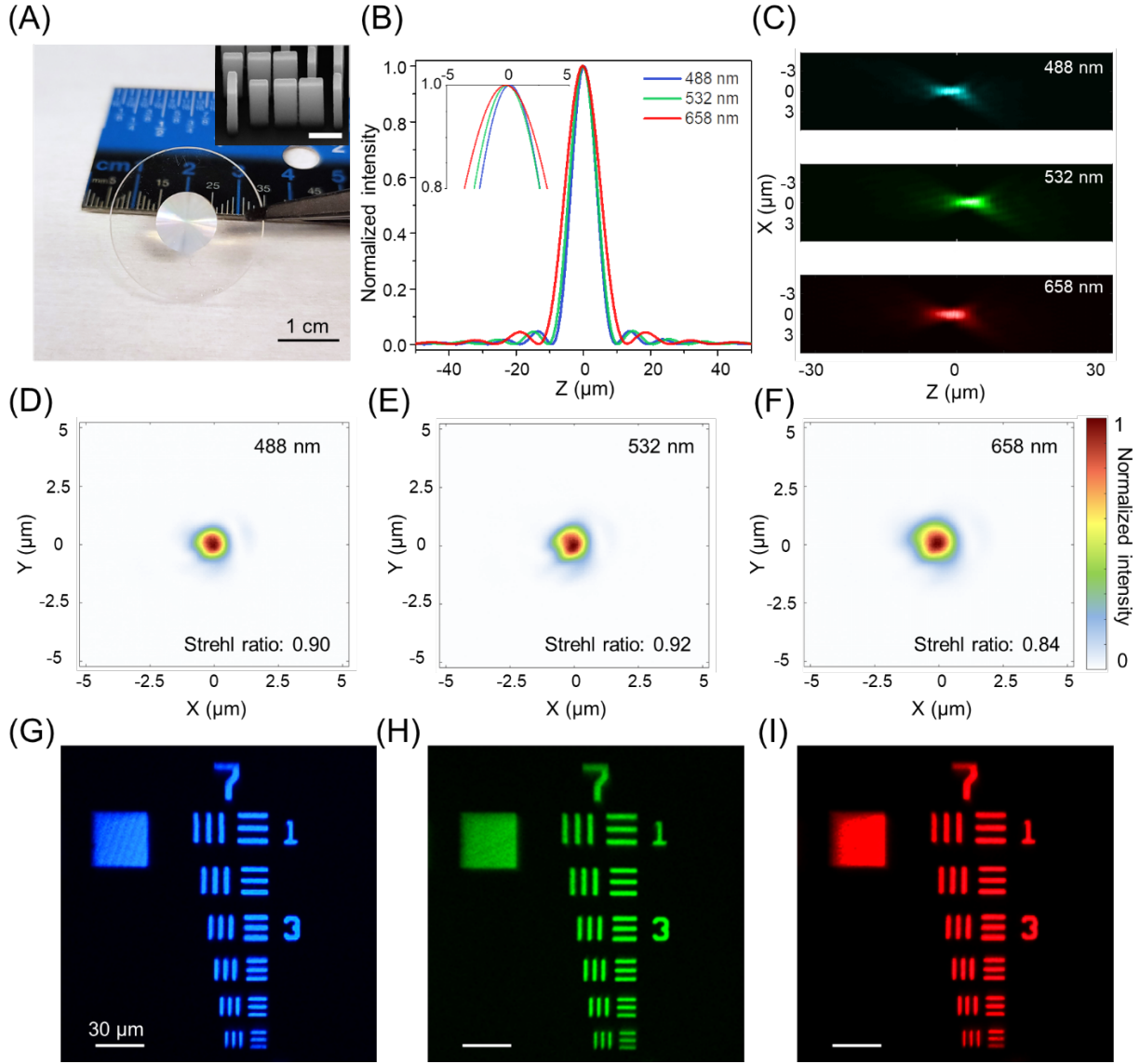

**Figure 4. 1-cm-diameter RGB-achromatic polarization-insensitive metalens (NA = 0.3)** (A), Photograph of the fabricated 1-cm-diameter metalens. The scale bar is 1 cm. The inset is the SEM image. The scale bar is 500 nm. (B), Simulations of focal intensity distribution along the optical axis at RGB design wavelengths. The inset is the zoomed-in view of the peaks. (C), Measured focal intensity distribution in the XZ plane at RGB wavelengths. (D)–(F), Measured normalized focal intensity distribution at the focal planes of $\lambda$ = 488 nm, 532 nm, and 658 nm, respectively. (G)-(I), Imaging results of the group#7 of the USAF resolution target under blue, green, and red-light illumination. The scale bars are 30 μm.

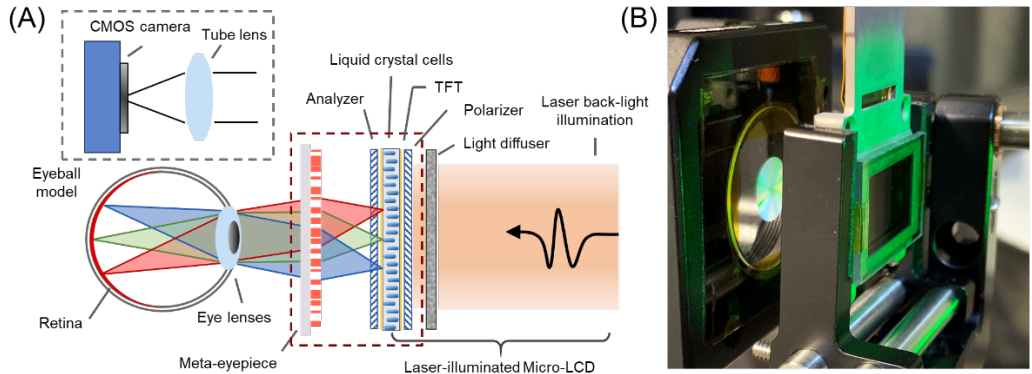
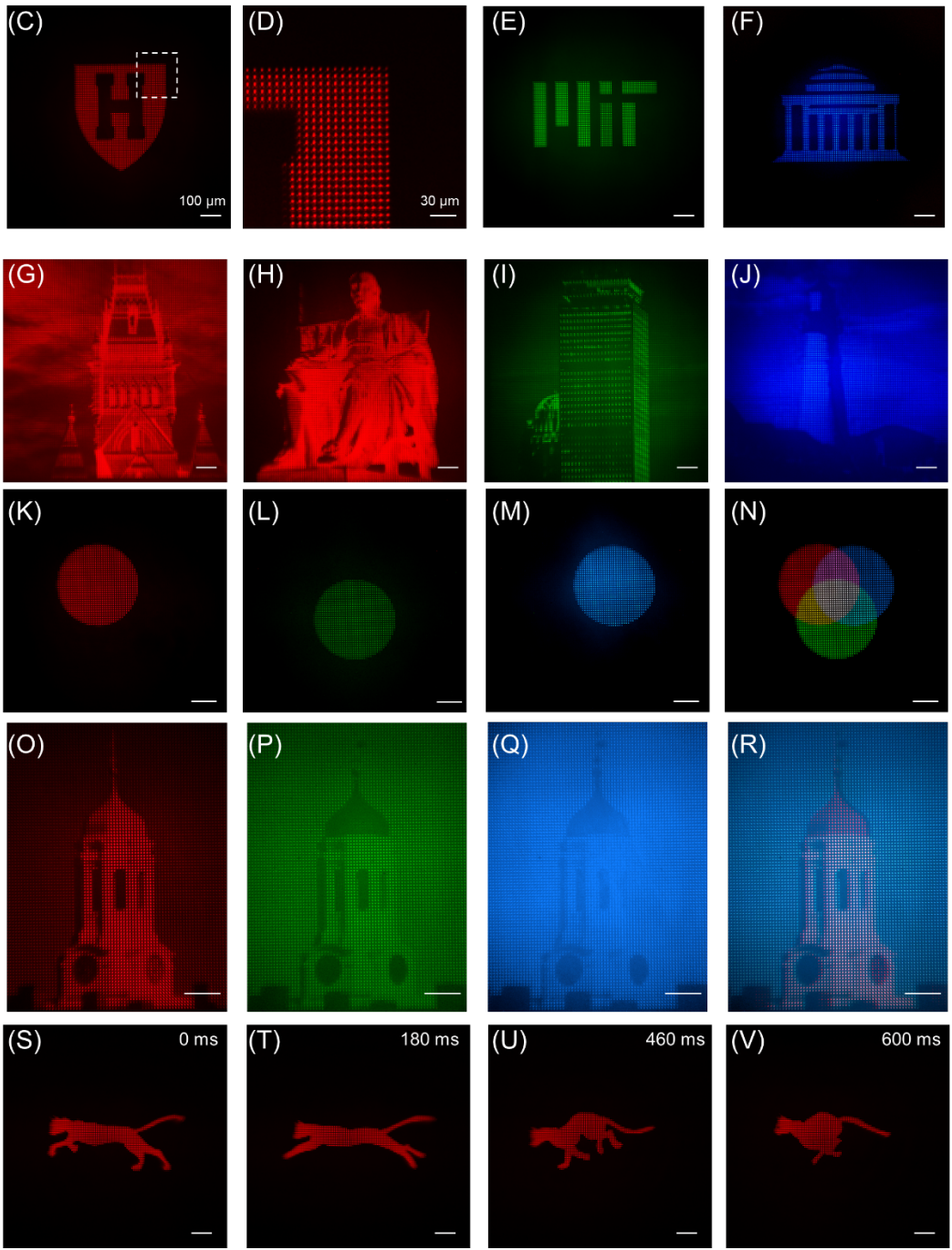

**Figure 5. Virtual-reality imaging demo using the meta-optics eyepiece.** (A), Schematic of the virtual-reality near-eye projection setup comprising an RGB-achromatic meta-optic eyepiece and a laser-illuminated micro-LCD. (B), Photography of the optical setup corresponding to the red dashed line in (A). The micro-LCD is mounted on a motorized stage and in front of the flat meta-optics. (C), Binary VR imaging result showing a Harvard logo in red color. The scale bar is 100 µm unless noted. (D), The zoomed-in view of the dash-lined area in (C). It shows the meta-optics can resolves every single pixel of the micro-LCD. The scale bar is 30 µm. (E)-(F), Binary imaging result of MIT logos in green and blue color, respectively. (G)-(H), Grey-scale VR imaging results showing a Harvard building and statue in red color. (I)-(J), Grey-scale VR imaging results of a Boston building and a light house in green and blue color, respectively. (K)-(M), VR images of a red, green, and blue circles. (N) Simulated VR image by superposing (K) to (M). (O)-(Q), Grey-scale VR imaging results of a Harvard tower in red, green, and blue channels, respectively. (R), Simulated full-color VR imaging result by combining RGB image channels shown in (O)-(Q). (S)-(V), A VR imaging movie at different frames showing a running cat. The movie can be found in the supplementary material. The near-eye display has a refresh rate of 60 Hz.